# LONG-TERM VARIATIONS OF GEOMAGNETIC ACTIVITY AND THEIR SOLAR SOURCES

**Kirov B.[1], Obridko V.N.[2], Georgieva K.[1], Nepomnyashtaya E.V.[2], Shelting B.D.[2]**

*1 – Space Research ant Technologies Institute - BAS, Sofia, Bulgaria*
*2 – IZMIRAN, Russia*

email: bkirov@space.bas.bg

## Abstract

Geomagnetic activity in each phase of the solar cycle consists of 3 parts: (1) a "floor" below which the geomagnetic activity cannot fall even in the absence of sunspots, related to moderate graduate commencement storms; (2) sunspot-related activity due to sudden commencement storms caused by coronal mass ejections; (3) graduate commencement storms due to high speed solar wind from solar coronal holes. We find that the changes in the "floor" depend on the global magnetic moment of the Sun, and on the other side, from the height of the "floor" we can judge about the amplitude of the sunspot cycle.

## 1. Introduction

As early as in 1852 it was noted that geomagnetic disturbances are related to solar activity [Sabine, 1852], and in 1982 it was found that geomagnetic activity is caused by two types of solar agents, the first one related to sunspots and caused by coronal mass ejections (CMEs), and the other one –not related to sunspots and caused by high speed solar wind from solar coronal holes [Feynman, 1982]. CME-caused geomagnetic disturbances have a maximum in sunspot maximum, and coronal holes-related disturbances – on the descending phase of sunspots. In the present study we aim to determine how these two components of geomagnetic activity vary from cycle to cycle, and the variations of which solar agents cause these changes.

## 2. Components of geomagnetic activity

Feynman [1982] noted that if a geomagnetic activity index (e.g. *aa*-index) is plotted as a function of the sunspot number *R* (Fig.1), all *aa*-index values lie above a line given by the equation $aa_R = a_0 + b.R$. Feynman [1982] suggested that $aa_R$ is the geomagnetic activity caused by sunspot-related solar activity, and the *aa* values above this line represent the geomagnetic activity caused by high speed solar wind: $aa_P = aa - aa_R$.

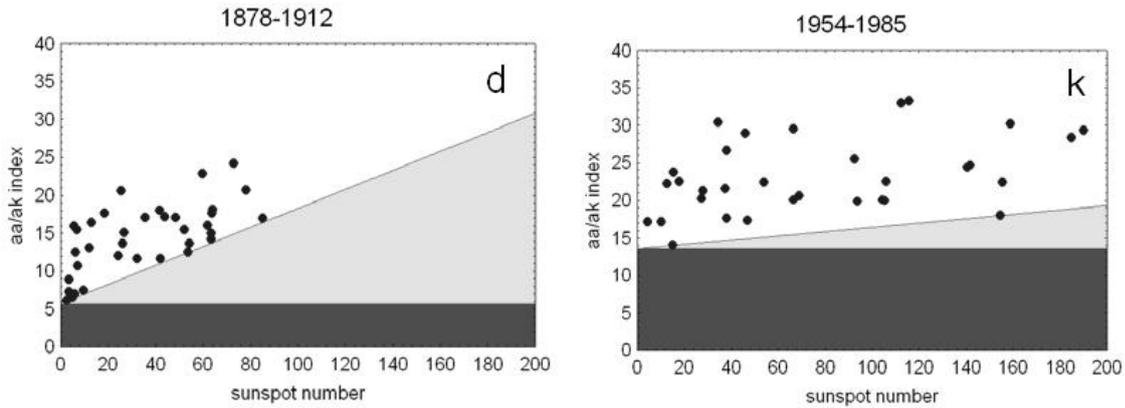

*Fig.1. Dependence of aa-index on R in the period 1878-1912 (left) and 1954-1985 (right)*

Feynman [1982] determined that $aa_R = 5.38 + 0.12 \cdot R$. We calculated the values of $a_0$ and $b$ in consecutive periods of around 30 years, each including 3 sunspot cycles (cycles 9-11, 10-12 and so on). The calculated values of the coefficients demonstrate a clear quasi-secular cycle (Fig. 2).

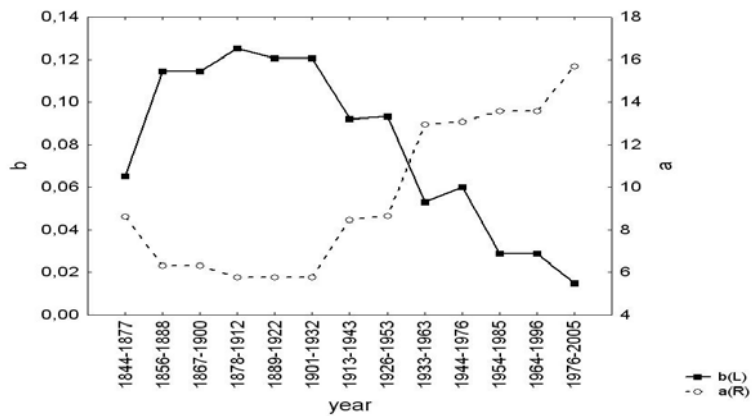

*Fig. 2. Cyclic variations of the coefficients $a_0$ and $b$*

Actually, geomagnetic activity can be divided into 3 rather than 2 components. The first one is the "floor", equal to the $a_0$ coefficient which represents the geomagnetic activity in the absence of sunspots. It is practically determined by the activity in the cycle minimum and varies smoothly from cycle to cycle. The second component is the geomagnetic activity caused by sunspot-related solar activity which is described by the straight line $aa_T = b \cdot R$ so that $aa_R = a_0 + aa_T$. The slope $b$ of this line also changes cyclically. The third component $aa_P$ (the value above $aa_R$) is caused by high speed solar wind. It can be seen that when the coefficient $a_0$ is big (high "floor" of the geomagnetic activity), the geomagnetic activity is almost independent on the sunspot number (small coefficient $b$), and when the "floor" is low, the geomagnetic activity quickly grows with growing sunspot number.

Fig.3 demonstrates that both the "floor" of the geomagnetic activity $a_0$ and the slope $b$ don't depend on the phase of the sunspot cycle (upward or downward branch), but are different in different intervals. Besides, the scatter of values above $aa_R$ in any interval is much bigger during the downward branch which has to be expected because of the strong influence of high speed solar wind streams on the geomagnetic activity in this phase of the sunspot cycle [Georgieva, K., Kirov, B, 2005].

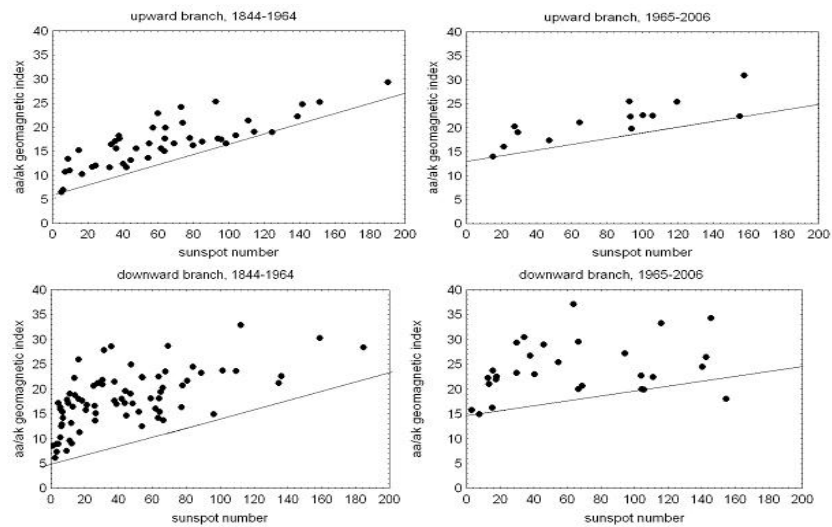

***Fig. 3.*** *Dependence of aa on R in the upward (upper panel) and in the downward branch (lower panel) of the sunspot cycle in the period 1844 – 1957 (left) and in the period 1955 – 1985 (right)*

The question now is which solar agents cause the nonzero values of the geomagnetic activity "floor", and which factors lead to the changes in the coefficients $a_0$ and $b$.

### 3. Types of geomagnetic storms

The geomagnetic storms differ in intensity as well as in characteristics. The intensity of the storm is determined by the parameters D, H и Z – the magnetic declination, change of the horizontal and vertical components of the Earth's magnetic field, respectively, measured in nT (Table 1), and the storm can be with a sudden or gradual commencement. In order to determine the origin of different storms, [Шельтинг, Обридко, 2011] compared the occurrence frequency of all, strong, moderate and weak storms with sudden and with gradual commencement to the number of sunspots. They found that the occurrence frequency of the sudden commencement storms correlates with the sunspot number, with correlation coefficient 0.872 +/- 0.06. In contrast, the correlation coefficient of the occurrence frequency of the gradual commencement storms and the sunspot number is practically zero (0.014 +/- 0.13), and their maximum is shifted by 1-3 years after the sunspot maximum. Further analysis shows that there is high correlation among storms of different intensity inside each of the groups, and absolutely no correlation between the two groups of storms. From there the authors make the conclusion that these two types of storms have different origin. Sudden commencement storms are caused by coronal mass ejections (CMEs). As the solar coronal mass ejections are associated with active areas, that is with local magnetic fields, a high correlation is observed between the occurrence frequency of sudden commencement storms and the number of sunspots.

Gradual commencement storms are caused by high speed solar wind. It originates from solar coronal holes whose maximum is on the downward branch of the sunspot cycle. This leads to two maxima in geomagnetic activity in the course of the sunspot cycle, the second one being due mainly to the occurrence of many gradual commencement storms [Tsurutani, et al. 2006].

## 4. Geomagnetic activity "floor"

The geomagnetic activity "floor" is the geomagnetic activity in the absence of any sunspots. Practically it is determined by the geomagnetic activity in the sunspot minimum. As seen in Fig.4, during the whole investigated period, the geomagnetic activity with *aa*-index between 10 and 30 is caused by gradual commencement storms, but this is especially well pronounced in sunspot minimum periods.

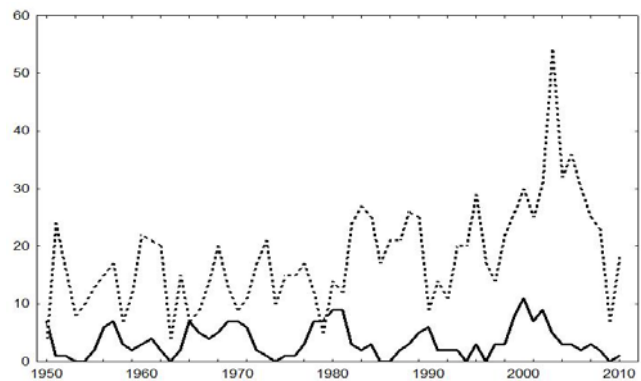

*Fig.4*. Annual number of weak sudden commencement storms (solid line) and gradual commencement storms (dotted line).

We can conclude that the geomagnetic activity "floor" is determined by non sunspot-related solar activity. Fig.5 demonstrates that the variations in the geomagnetic activity floor follow, with the exception of cycle 13, the variations in the number of 30hour intervals with *aa*-index between 10 and 30 in the minimum of the respective cycle. Moreover, from cycle 14 to 21 (around 1985), an increase is observed in both the number of 3-hour intervals with $10<aa<30$, and of the geomagnetic activity floor, after which they both begin decreasing. The correlation between the two quantities is 0.85 with $p<0.01$.

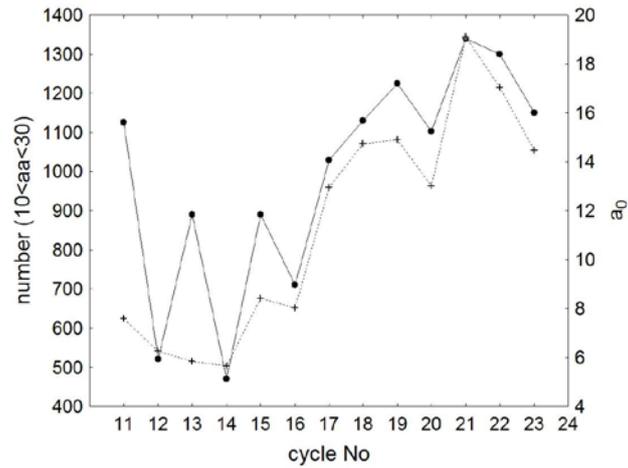

*Fig.5. Number of 3-hour intervals with aa between 10 and 30 (solid line) and the geomagnetic activity floor $a_0$ (dotted line), in consecutive sunspot cycle minima.*

It is seen that the geomagnetic activity floor in each cycle is determined by the geomagnetic activity in the range 10<aa<30 in the cycle minimum, and as far as this geomagnetic activity, especially in the cycle minimum, is related as shown above to gradual commencement storms, the floor is determined by the high speed solar wind streams causing gradual commencement storms. Therefore, the reason for the changing geomagnetic activity floor is the changing high speed solar wind reaching the Earth during sunspot minimum.

It is interesting to note that at the same time as the increase in the floor changed to decrease (cycle 21), the increase in the solar global magnetic moment changed to a sharp decrease[Обридко, Шельтинг, 2009] - (Fig.6). It seems that the change in the solar global magnetic moment is reflected in the geomagnetic activity caused by the high speed solar wind during the sunspot minimum through its effects on the thickness of the heliospheric current sheet [Simon and Legrand, 1987].

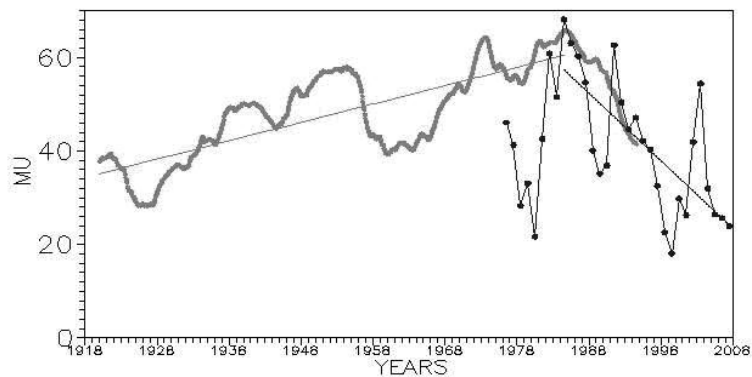

*Fig.6. Evolution of the solar magnetic moment from 1918 to 2006.*

## 5. Coefficient b (slope)

The sunspot-related geoeffective solar agents are the CMEs. It is known that the geoeffectiveness of CMEs vary little in the course of the sunspot cycle; considerably changes the geoeffectiveness of magnetic clouds (a subclass of CMEs with high and smoothly rotating magnetic field), but their number is negligible compared to the total number of CMEs except around sunspot minimum when their geoeffectiveness is low. Therefore, the varying contribution of CMEs to geomagnetic activity from year to year depends on their varying number. [Georgieva and Kirov, 2005]. Fig.7 presents the ratio of CMEs to the number of sunspots for the period 1996-2012. For the number of CMEs, the SOHO/LASCO CME catalog is used (http://cdaw.gsfc.nasa.gov/CME_list/) as it provides a fairly homogenous data set.

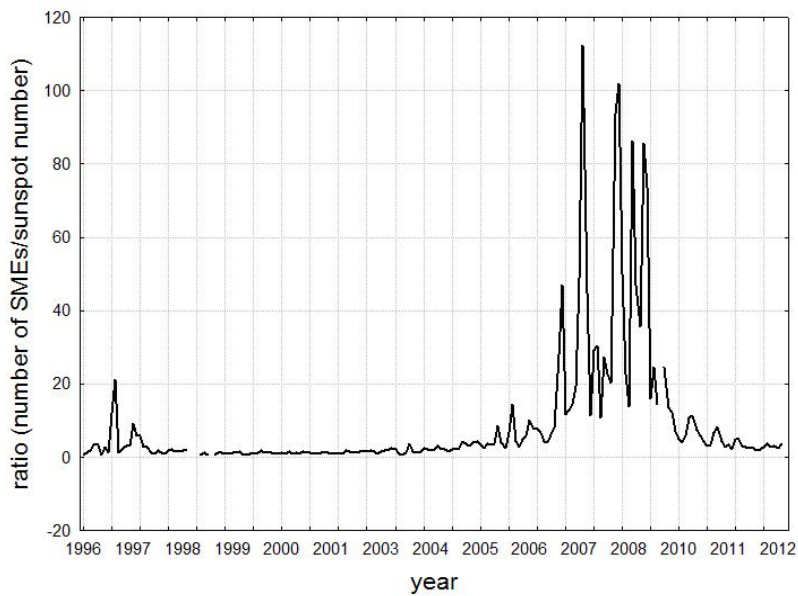

*Fig. 7. Ratio of the number of CMEs to the sunspot number (monthly averages).*

It can be seen that the ratio is almost constant during the greater part of the sunspot cycle, with the exception of the periods of sunspot minima, especially in the minimum between cycles 23 and 24 which is known to be very peculiar. In 2007, 2008, and 2009 we had total annuals of 20, 70, and 60 CMEs, respectively, with the average annual number of sunspots between 3 and 7. A small change in the correlation between the sunspot number and the number of CMEs is observed in the previous sunspot minimum also, but it is only due to CMEs in a few single months. We can conclude that the increasing geomagnetic activity with increasing sunspot number (the coefficient *b*) is due to the increasing number of CMEs with increasing number of sunspots.

## 6. Forecasting the following cycle

Fig.8 demonstrates that the sunspot number in the cycle maximum is related to the geomagnetic activity floor in the same cycle. The correlation between the two variables is 0,792 with p=0.001.

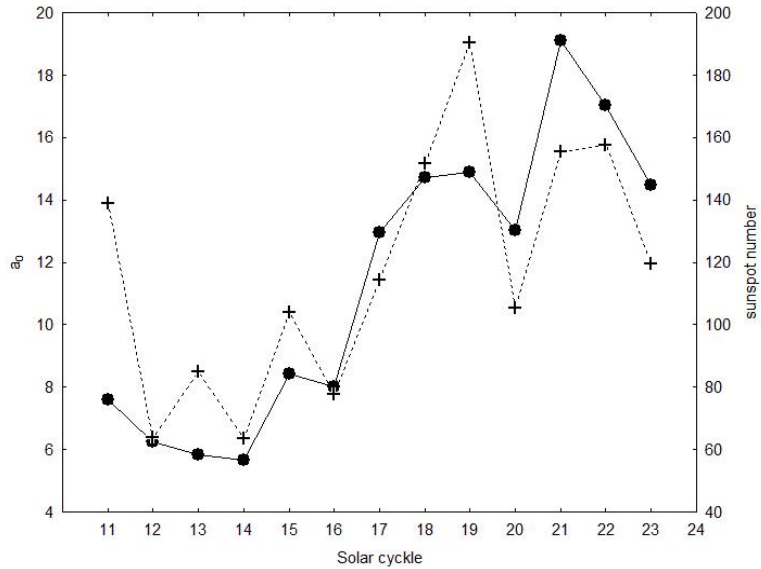

*Fig.8. Geomagnetic activity floor (solid line) and sunspot number in the cycle maximum (dotted line)*

From this it follows that the characteristics of the whole cycle are set already at its beginning, and measuring the geomagnetic activity we can forecast to a great extend its maximum. On the other hand, given that the evolution of the Sun's magnetic moment is indicative of the changes in $a_0$, we can evaluate in advance the direction of change of the geomagnetic activity "floor".

## 6. Summary and conclusions

The geomagnetic activity consists of three components: (1) $a_0$ – the geomagnetic activity "floor", theoretically equal to the activity at zero number of sunspots. Practically this activity is determined by the gradual commencement geomagnetic disturbances with $10<aa<30$ in the beginning of the sunspot cycle; (2) $aa_T$ – the geomagnetic activity caused by CMEs whose number linearly increases with increasing number of sunspots $aa_T = b.R$, so that $aa_R = a_0 + b.R$; (3) $aa_P$ – the values of $aa$ above $aa_R$, caused by high speed solar wind streams from solar coronal holes. We have found that the variations of $a_0$ have a quasi-secular cycle, with the direction of variation following the variation of the global magnetic moment of the Sun. On the other hand, we have found that $a_0$ determined by the geomagnetic activity in the beginning of a cycle is directly related to the sunspot maximum in the same cycle. Therefore, if we know the direction of change (increasing or decreasing) of the global magnetic moment, we can forecast the increase or decrease of $a_0$ in the next cycle, and from there – the increase or decrease of the sunspot maximum. Moreover, during the sunspot minimum we can quite accurately forecast the following sunspot minimum.

**Table 1**

| **Storm intensity** | **D** | **H** | **Z** |
|---|---|---|---|
| Minor | 100-139 | 80-125 | 40-90 |
| Moderate | 140-200 | 126-200 | 31-140 |
| Strong | 201-290 | 201-270 | 141-250 |
| Severe | $\geq$291 | $\geq$271 | $\geq$251 |